\documentclass[aps,pre,psfig,twocolumn,showpacs,superscriptaddress,reprint,floatfix]{revtex4-2}
\usepackage{amsmath}
\usepackage{mathtools}
\usepackage{amssymb}
\usepackage{color}
\usepackage{graphics}
\usepackage{array}
\usepackage{epsfig}
\usepackage[normalem]{ulem} 
\usepackage{cancel}
\usepackage[mathlines]{lineno}
\graphicspath{{Figures/}}
\usepackage{svg}
\usepackage{xtab,afterpage,longtable}
\usepackage{booktabs}   
\usepackage{ltablex}
\usepackage{enumitem}
\relax
\usepackage{float}
\usepackage{orcidlink}
\usepackage{hyperref}
\hypersetup{colorlinks=true,linkcolor=blue,urlcolor=blue,citecolor=blue}

\begin{document}
\title{Two-Stream Instability and Bernstein-Greene-Kruskal Mode Formation in Coulomb One Component Plasma}
\author{Ajaz Mir\orcidlink{0000-0001-5540-4967}}
\email{ajazmir.physics@gmail.com}
\affiliation{University of Kashmir, North Campus, Delina, Baramulla 193103, Jammu \& Kashmir, India}
\author{Rauoof Wani\orcidlink{0000-0003-0467-2691}}
\author{Sanat Tiwari\orcidlink{0000-0002-6346-9952}}
\email{sanat.tiwari@iitjammu.ac.in}
\affiliation{Indian Institute of Technology Jammu, Jammu  181221, Jammu \& Kashmir, India}
\author{Abhijit Sen\orcidlink{0000-0001-9878-4330}}
\affiliation{Institute for Plasma Research, Gandhinagar 382428,  Gujarat,  India}
\affiliation{Homi Bhabha National Institute, Anushaktinagar, Mumbai 400094, India}
\date{\today}
\begin{abstract}
\textcolor{black}{
We investigate the Two-Stream Instability in a strongly coupled plasma using classical molecular dynamics simulations with long-range Coulomb interactions between particles. The nonlinear evolution of the instability is identified by the emergence of a Bernstein–Greene–Kruskal (BGK) mode.}
Our simulations capture key microscopic effects, such as inter-particle correlations, collisional dynamics, and coherent wave-particle interactions—features often absent in traditional fluid and kinetic models, including Particle-In-Cell and Vlasov approaches.
In the linear regime, the instability grows rapidly and saturates within a few tens of plasma periods. As the system transitions into the nonlinear saturation phase, a single BGK mode emerges. This mode (or phase-space hole) becomes dynamically unstable in the nonlinear regime, characterized by a continuous decay of electrostatic energy over time. An energy budget analysis reveals a bump in an otherwise thermal spectrum, indicating the excitation of a coherent mode, further confirmed through a numerical rendering of the dispersion relation. The pairwise interaction plays a crucial role: pronounced instability and BGK mode formation occur with long-range Coulomb forces, while such structures are suppressed under shielded Coulomb interactions. We observe the emergence of a single BGK mode across all coupling strengths in the fluid regime, provided the streaming velocity exceeds a critical threshold.  
\end{abstract}
\maketitle
\section{Introduction}
\label{intro}
\paragraph*{}
The Two-Stream Instability (TSI) arises in a plasma system with two interpenetrating beams with velocity distribution departing from the Maxwellian~\cite{Krall_MGH_1973}. This electrostatic instability grows rapidly due to bunching and charge separation in real space. It is a mechanism that transfers energy from particles to waves, an inverse Landau damping problem~\cite{Morgado_PRL_2017, Landau_JP_1946}. For a typical plasma with kinetic energy dominating the average interaction energy, fluid and kinetic models such as Particle-In-Cell (PIC) and Vlasov simulations are appropriate to study the TSI~\cite{Krall_MGH_1973, Buneman_PRL_1958, Ng_PRL_2005, Hutchinson_POP_2017, Che_PNAS_2017, Dieckmann_PRL_2005}. 
In this paper, we focus on a class of plasmas whose average interaction energy is comparable to or higher than the kinetic energy of particles - the so-called strongly coupled plasmas (SCPs). For such systems, the correlations are important, and large-angle collisions significantly contribute to the collective dynamics. 
To capture these effects, we employ classical molecular dynamics (MD) simulations to investigate the evolution of the Two-Stream Instability in a one-component plasma (OCP) system, with particular emphasis on the formation and stability of a single Bernstein–Greene–Kruskal (BGK) mode. Unlike conventional kinetic approaches, the MD framework inherently incorporates correlations, collisional dynamics, viscosity, and wave–particle interactions. 
\paragraph*{}
The electrostatic wave associated with the Two-Stream Instability extracts energy from the kinetic motion of drifting particles in the two beams, analogous to the mechanism of the Buneman instability~\cite{Buneman_PRL_1958}. During the nonlinear evolution of the TSI, a coherent phase-space vortex — commonly referred to as a BGK mode — emerges, signaling localized depletion or trapping of charged particles~\cite{BGK_PR_1957, Ng_PRL_2005, Hutchinson_POP_2017}. The formation of such structures offers critical insights into the nonlinear saturation mechanisms of various electrostatic plasma instabilities. The BGK modes represent exact, time-independent nonlinear solutions of the Vlasov–Poisson system in collisionless plasmas~\cite{Ng_PRL_2005, Hutchinson_PRL_2018, Hutchinson_POP_2017}. In weakly collisional regimes, a single BGK vortex can be stable and persist over long timescales. These phase-space structures are often interpreted as the final saturated state of an instability, in which particle trapping balances wave growth within a self-consistent potential well. 
\textcolor{black}{In present simulations, we observe that the sustained decay in electrostatic energy arises due to the instability of the BGK mode~\cite{Morse_PRL_1969, Morse_POF_1969}.}
\paragraph*{}
\textcolor{black}{The existence of BGK modes and the Two-Stream Instability are well established in plasmas, however, the present study extends these concepts to the strongly coupled regime using fully kinetic molecular dynamics simulations.
Due to the visco-elastic nature of SCPs, such plasmas cannot be adequately modeled using the fluid-Maxwell framework. Similarly, Vlasov and PIC simulations, rooted in weakly coupled assumptions, fail to capture the enhanced collisionality and its effects. The study reveals the high sensitivity of TSI to the pairwise interaction potential. The instability manifests clearly with long-range Coulomb interactions but vanishes even with minimal shielding.
Moreover, the present work showed that the TSI can be found with isotropic and anisotropic geometries, and also showed that a threshold transverse domain length is required to sustain the BGK mode formation.}
\paragraph*{}
The development of phase-space vortices and formation of whirling structures are ubiquitous in plasma systems that are of enormous fundamental interest and also find wide-ranging applications~\cite{Hutchinson_RMP_2024}. The TSI is the basic mechanism of two-stream free electron lasers~\cite{McNeil_PRE_2004, Freund_CH_1992} as well as a means of generating Terahertz radiation in a plasma source~\cite{Liu_IEEE_2008, Ignat_PRA_1970}. It is directly relevant to plasma turbulent heating~\cite{Jensen_PRL_1967,Thode_PRL_1973,Davidson_PRL_1970}, and irregularities in the ionosphere~\cite{Farley_PRL_1963}. The instability also drives the Langmuir collapse in the solar corona~\cite{Che_PNAS_2017} and affects ion and electron transport in presheaths~\cite{Baalrud_PSST_2016}. The TSI has been studied in quantum and classical plasmas to understand the generation of electromagnetic radiation and magnetic fields~\cite{Liang_PRE_2021, Hu_PRE_2022}. The TSI has been investigated in Bose-Einstein condensates~\cite{Tercas_PRA_2009}, Fermi-Dirac plasmas~\citep{Akbari_APSC_2016}, superfluids~\citep{Schmitt_PRL_2014}, graphene~\citep{Aryal_PRE_2016, Morgado_PRL_2017} and solid-state plasmas~\citep{Hu_PRB_1991}. Typically, the TSI has been obtained in a plasma by creating two bumps in a Maxwellian velocity distribution~\cite{Ghorbanalilu_LPB_2014}.  
\paragraph*{}
The TSI is inherently a kinetic instability, due to anisotropic or non-Maxwellian velocity distributions, and cannot be appropriately treated using a fluid description. This is primarily because plasmas, in realistic settings, often deviate from thermal equilibrium and exhibit velocity distributions that are non-Maxwellian. Therefore, the kinetic effects play a crucial role in the evolution of the instability. Consequently, it has been predominantly investigated using PIC simulations~\cite{Ghorbanalilu_LPB_2014, Che_PNAS_2017, Cruz_PRE_2021, Dieckmann_PRL_2005, Sakawa_PRE_2021}, which, however, omit kinetics effects occurring due to strong correlations. Fully kinetic particle-based molecular dynamics simulations are a robust tool to capture the kinetic effects of fluid instabilities. Various instabilities studied using MD simulations are Kelvin-Helmholtz instability~\cite{Ashwin_PRL_2010, Sanat_JPP_2014}, Rayleigh-Taylor instability~\cite{Kadau_PNAS_2004, Rauoof_SR_2022,Rauoof_POP_2024}, Rayleigh-B\'{e}nard convection~\citep{Rapaport_PRL_1988, Charan_POP_2015, Pawandeep_PRE_2019} and Bump-On-Tail (BOT) instability~\citep{Williams_PRR_2019}.
\textcolor{black}{Recently, Williams~\textit{et al.}~\cite{Williams_PRR_2019} provided a comprehensive study of the BOT instability across coupling and interaction regimes in a long-range Coulomb system. Our effort marks only the second MD-based examination of TSI in strongly coupled plasmas, following the pioneering work by Williams~\textit{et al.}~\cite{Williams_PRR_2019}.} The present study offers significant insight into the growth and evolution of the Two-Stream Instability, emphasizing the roles of system dimensionality, coupling strength, and shielding effects. These factors critically influence the development of the instability and have received comparatively less attention in prior investigations.
\paragraph*{}
Strongly coupled plasmas—where the Coulomb interaction energy dominates over thermal energy—are commonly modeled as one-component plasmas (OCPs), encompassing both pure Coulomb and screened Coulomb (Yukawa) systems~\citep{Ichimaru_RMP_1982}. In thermodynamic equilibrium, frictionless pure Coulomb OCPs (COPCs) are characterized by the Coulomb coupling strength $\Gamma$, defined as the ratio of the average Coulomb potential energy $\left\langle E_C \right\rangle$ to the average thermal kinetic energy $\left\langle E_T \right\rangle$, expressed as:
\begin{equation}
\label{Eqn_1}
\Gamma = \frac{\left\langle E_C \right\rangle}{\left\langle E_T \right\rangle} 
       = \frac{Q^2}{4 \pi \epsilon_0 a} \frac{1}{k_B T}
\end{equation}
where $a = [3/(4\pi n)]^{1/3}$ is the average inter-particle separation or Wigner-Seitz radius, and $n$ is the number density of dust particles. $Q = -Z e$ is the particle charge, and $Z$ is  the number of elementary charges of magnitude $e$ residing on the charged particle. $\epsilon_0$ and $ k_B$ are the permittivity of free space and Boltzmann constant, respectively. T is the particle kinetic temperature related to its thermal velocity $v_{th}$ given by 
\begin{equation}
\label{Eqn_2}
    v_{th} =\ \sqrt{ \frac{3 k_B T}{m}} =\ \sqrt{\frac{3}{m} \frac{Q^2}{4\pi \epsilon_0 a} \frac{1}{\Gamma} }
\end{equation}
where $m$ is the mass of the particle.
\paragraph*{}
The manuscript is organized into two sections. 
The performed classical MD simulation are explained in section~\ref{MD_model}. The signatures of TSI are explained through the formation of the phase-space vortex and real-space bunching. The mechanism of nonlinear saturation of TSI is explained through the formation of the BGK mode. \textcolor{black}{A detailed spectral analysis is also performed to understand the thermalization of the instability.} A summary of the work with conclusive remarks and a discussion on the future extension of the present work are presented in section~\ref{Sum_Con}.
\section{Molecular dynamics simulations}
\label{MD_model}
\paragraph*{}
We have conducted three-dimensional classical molecular dynamics simulations to investigate the TSI in a strongly coupled one-component plasma. The dynamics of each charged particle in the ensemble are governed by Newton’s equations of motion, expressed by
\begin{equation}
    \label{Eqn_4}
    m \ddot{\textbf{r}}_i  =\ - \nabla \sum \phi_{ij} .
\end{equation}
The charged particles interact via a pair-wise Coulomb potential given by
\begin{equation}
 \label{Eqn_5}
 \phi_{ij} = \frac{Q^2}{4\pi \epsilon_0 r_{ij}} 
\end{equation}
where $r_{ij}$ is the distance between the $i^{th}$ and $j^{th}$ particles. We generate the trajectories $\textbf{r}_i (t)$ for all particles by integrating the equations of motion (\textit{i.e.,} Eq.~\eqref{Eqn_4}) for each particle in the ensemble. The open-source Large-scale Atomic/Molecular Massively Parallel Simulator (LAMMPS)~\cite{LAMMPS_2022} code has been used for the present MD simulations. The long-range Coulomb potential was used with particle-particle-particle-mesh (PPPM) algorithm~\citep{David_PRE_2003}, first described by Hockney and Eastwood~\citep{Hockney_TF_1988}. We also employed the standard Ewald summation for Coulomb OCP, which takes into consideration the neutralizing background~\cite{Deserno_JCP_1998}. 
\paragraph*{}
To begin with, we distributed the point-charged particles homogeneously inside a cubic box ($L_x = L_y = L_z$) with periodic boundary conditions in all directions. The particle parameters are tabulated in Table~\ref{Table_1}.
\begin{table}[!ht]
\caption{Simulation parameters for TSI.}
{\renewcommand{\arraystretch}{2.5}
\begin{tabular}{|l|l|}
\hline  
\textbf{Particle parameter}          & \shortstack{\\ \textbf{Cubic configuration}  
                                        \\ $(L_x = L_y = L_z)$}          \\ \hline 
\textbf{No. of particles,    $N$}    & $4 \times 10^4$                   \\ \hline 
\textbf{Number density,      $n$}    & $2.923 \times 10^9$ m$^{-3}$      \\ \hline 
\textbf{Charge of particle,  $Q$}    & \shortstack{\\ $15\times 10^3$ e 
                                      \\ \small{e = Electron charge} }    \\ \hline 
\textbf{Mass of particle,    $m$}    & $6.9 \times  10^{-13}$  kg         \\ \hline 
\end{tabular}
}
\label{Table_1}
\end{table}
Further, we gave random velocities to the particles corresponding to a temperature T. To study a dynamical system at a particular coupling strength $\Gamma$ (corresponding to temperature T), we allow it evolve while in contact with a Nos\'{e}-Hoover~\citep{Nose_MP_1984, Hoover_PRA_1985} thermostat. We found that 400 $\omega_{pd}^{-1}$ is a reasonable time for the system to attain equilibrium around this temperature, \textcolor{black}{where $\omega_{pd}$ is the dust plasma frequency given by
$ \omega_{pd} = (n Q^2 / \epsilon_0 m)^{1/2}$.} As a next step, we let the system evolve freely for another 400 $\omega_{pd}^{-1}$ after detaching it from the thermostat.  During this phase of evolution, the system obeyed energy conservation, and its equilibrium was maintained around the desired temperature T. We found the total energy to be conserved up to $10^{-2} \%$. We also validated the equilibrium properties of the system by comparing with values available in the literature~\citep{Ott_CPP_2015}.
\begin{figure*}[ht!]
    \centering
    \includegraphics[width= \textwidth, height=0.65\textwidth]{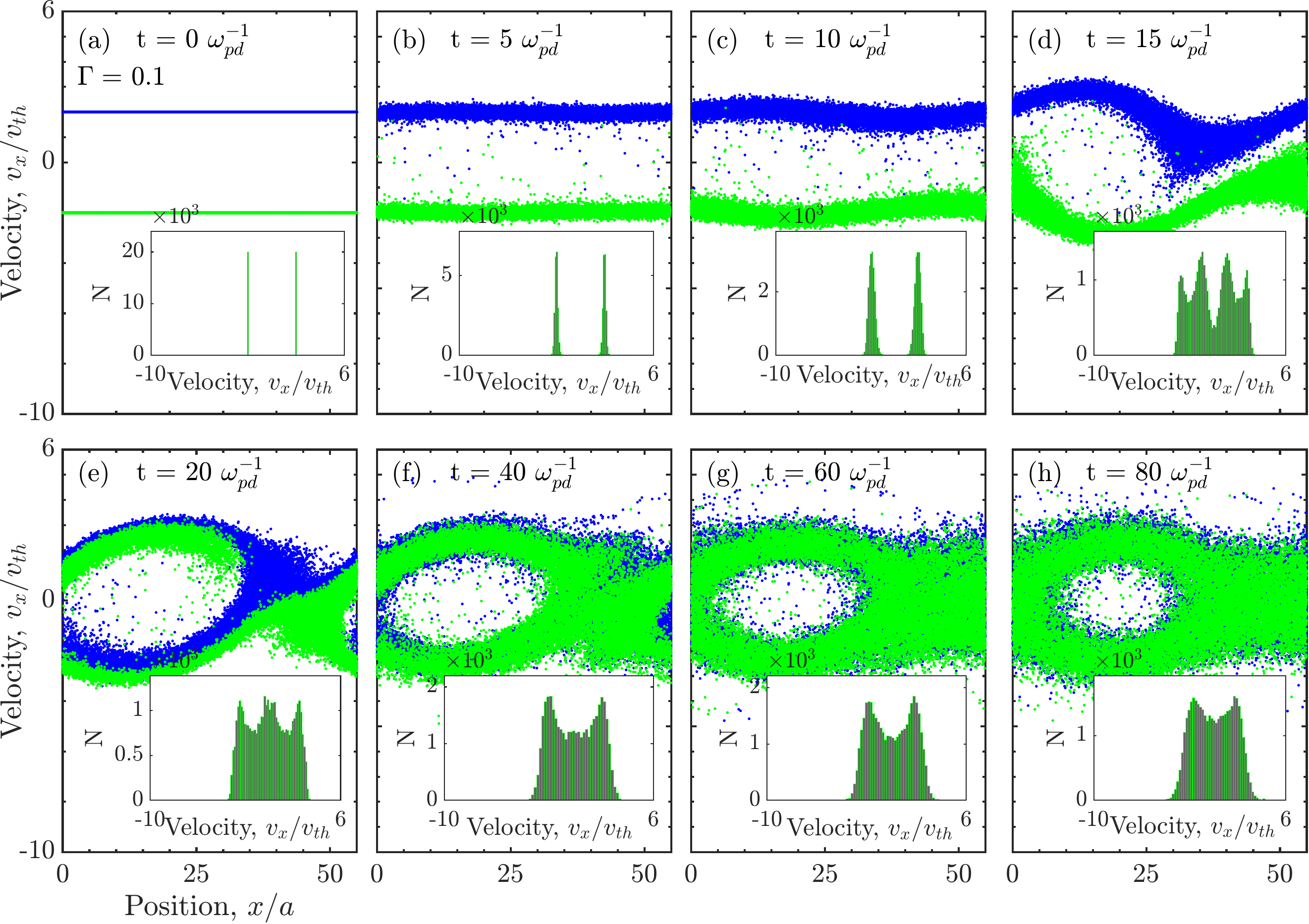}
\caption{
Phase-space evolution of the TSI for $\Gamma = 0.1$ in a long-range Coulomb OCP. Each dot represents a single negatively charged particle. Insets display the corresponding velocity distribution at each time snapshot. A full animation of the molecular dynamics simulation, showing the emergence of a BGK mode excited by TSI in Coulomb OCP, is available at the provided URL.
}
    \label{Figure_1}
\end{figure*}
\begin{figure*}[ht!]
    \centering
    \includegraphics[width= \textwidth]{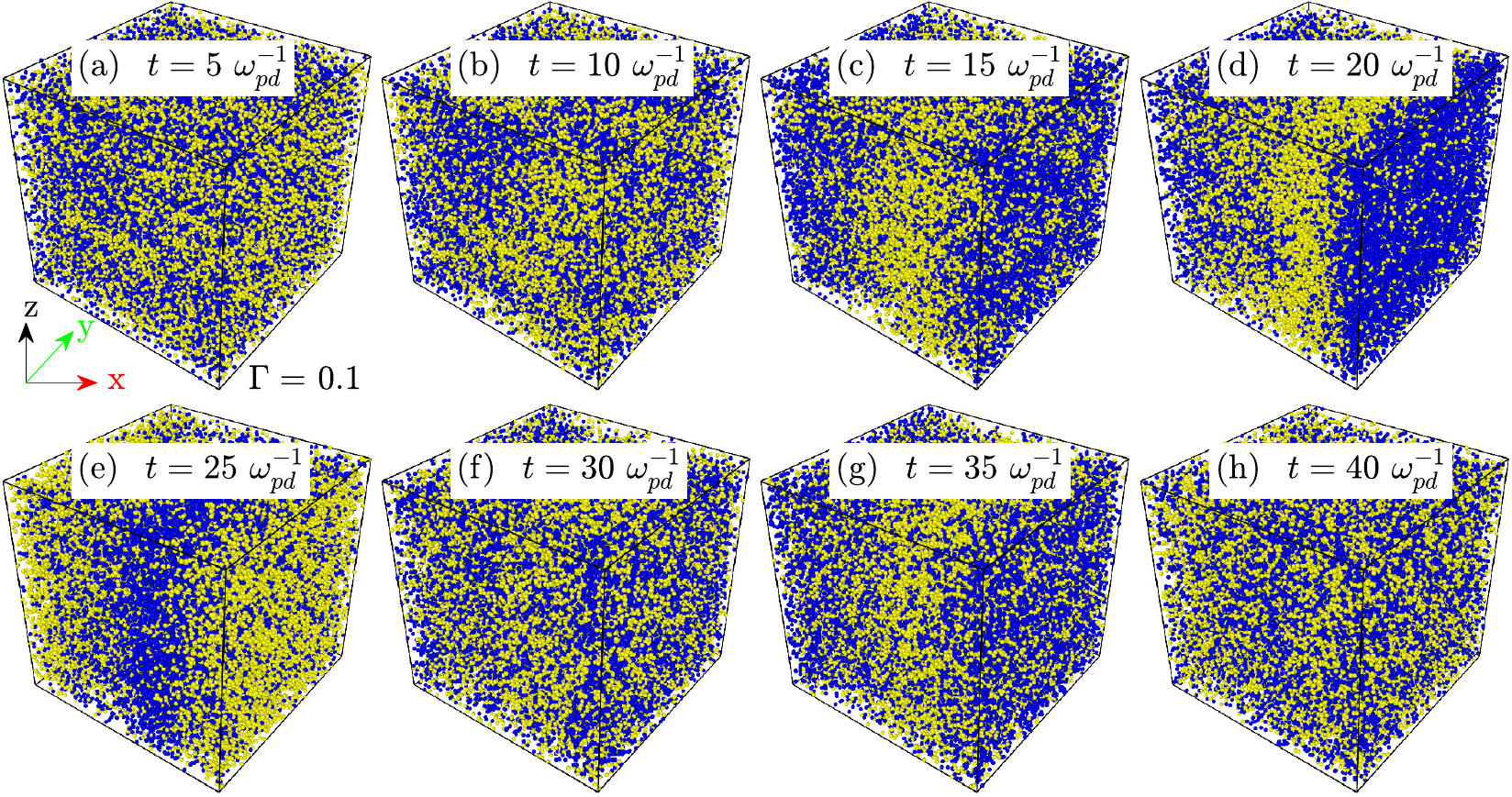}
    \caption{
    Real-space evolution of two interpenetrating particle streams (blue and yellow) in the MD simulation for $\Gamma = 0.1$. The streaming particles exhibit space-charge bunching at $t = 15,\ 20,\ \text{and}\ 25\ \omega_{pd}^{-1}$
    }
    \label{Figure_2}
\end{figure*}
\paragraph*{}
At this stage, we divided up the total particles in our system into two groups of equal number of particles to form two inter-penetrating velocity beams that are well separated in velocity-space (Dirac-delta velocity-distribution profiles) and allowed the system to evolve naturally. We have given velocity streaming in the x-direction, and the perturbation velocity is given by
\begin{equation}
    \label{Pert_eqn}
    v_{x}^{P} = v_x \pm v_0[ 1 + A\cos(k x) ]; \hspace{0.5cm} v_0 = \alpha v_{th}, 
\end{equation}
 where $v_0$ is the streaming velocity with strength $\alpha$. \textcolor{black}{The streaming velocity $v_0$ between the two interpenetrating beams is normalized by the thermal velocity $v_{th}$. A TSI is observed to emerge only when $v_0$ is near a threshold value, a behavior that is consistently seen across all the coupling strengths.} $A$  and  $k = m k_{m}$ are the amplitude and wave vector of perturbation, respectively. $m = 1, 2, 3,....$ is the mode of perturbation and $k_{m} =  2\pi / L_x$ is the minimum wave vector associated with a system of length $L_x$.
 \textcolor{black}{
A sinusoidal perturbation has been provided along with the streaming velocity to assist in the early growth of the instability. It also helps to excite two-stream instability at different mode numbers by assisting them initially. However, we clarify that, in this first principle approach, the instability of the maximally growing mode will appear even if no external perturbation is applied. 
 }
 \paragraph*{}
 While the majority of our results focus on pure long-range Coulomb systems, we have also examined Yukawa systems and found that the TSI is effectively suppressed even with very weak screening, at a Debye screening parameter as low as $\kappa = a/\lambda_D = 0.01$, where $\lambda_D$ is the Debye wavelength of the dusty plasma. This highlights the crucial role of long-range interactions in driving the kinetic instability.
\subsection{TSI in long-range Coulomb systems}
\label{TSI_MD}
\paragraph*{}
In our molecular dynamics simulations, the Two-Stream Instability manifests distinctly at a Coulomb coupling parameter of $\Gamma = 0.1$. The instability triggers the formation of coherent phase-space holes—localized regions where particles become trapped within the deep potential wells created by finite-amplitude electrostatic plasma waves. These phase-space holes, also known as Bernstein-Greene-Kruskal modes, serve as hallmark nonlinear structures representing the saturation stage of the instability. Their periodic formation acts as a natural nonlinear saturation mechanism, effectively limiting the growth of large-amplitude electrostatic waves generated by the TSI, consistent with prior theoretical and simulation studies~\cite{Eliasson_PR_2006, Hu_PRE_2022, Morse_POF_1969}. 
\paragraph*{}  
Figure~\ref{Figure_1} presents the phase-space evolution of an OCP with long-range Coulomb interactions. The initial conditions correspond to a velocity streaming strength of $\alpha = 2$ and a perturbation amplitude of $A = 0$. The onset of the Two-Stream Instability is evident in the early phase-space dynamics, characterized by the emergence of a void, resulting from charge bunching in real space. The formation of a single, coherent, and long-lived phase-space vortex that is clearly visible at $t = 20\ \omega_{pd}^{-1}$ confirms the development of the Two-Stream Instability and the associated BGK mode. By $t = 40\ \omega_{pd}^{-1}$, the two beams have significantly mixed, yet the remnants of the BGK structure remain discernible even at $t = 80\ \omega_{pd}^{-1}$. This persistence further supports the interpretation of the nonlinear saturation mechanism as BGK-mode formation. Insets in each panel display the corresponding velocity distribution, showing two distinct peaks at early times—indicative of the streaming beams. As the instability saturates, the distribution evolves towards a Maxwellian, reflecting the system’s approach to a new quasi-equilibrium state.
\paragraph*{}
The sequence of snapshots in Fig.~\ref{Figure_1} illustrates the formation and evolution of the Bernstein–Greene–Kruskal mode during the development of the TSI. Panel (a) shows the initial configuration consisting of two counter-streaming beams with equal and opposite Dirac-delta velocity distributions. Panel (b) depicts the linear growth phase of the instability, while panel (c) marks the termination of linear amplification. Beyond this point, nonlinear interactions begin to dominate, causing the particle beams to twist and distort in phase space. In panel (d), particles with velocities close to the phase velocity of the wave (which is zero in the wave frame) become trapped and begin to oscillate within the self-consistent potential well. As the system evolves through several bounce periods, a coherent eddy structure forms, as shown in panels (f), (g), and (h), indicating the emergence of a near-stationary BGK mode in the wave frame. The formation of this single, stable phase-space vortex highlights the role of particle trapping in the kinetic nonlinear saturation mechanism of the TSI.
\paragraph*{}
The hallmark of TSI is also manifested explicitly through the formation of space-charge bunches in real space, resulting from the mutual reinforcement of charged particles between the interpenetrating beams. This bunching is driven by the excess kinetic energy of the two streams. Figure~\ref{Figure_2} illustrates the evolution of real-space charge density in a three-dimensional OCP with long-range Coulomb interactions at a coupling strength of $\Gamma = 0.1$. Snapshots are shown from $t = 0$ to $t = 40\ \omega_{pd}^{-1}$, in increments of $5\ \omega_{pd}^{-1}$. Pronounced charge bunching appears between $t = 15$ and $t = 25\ \omega_{pd}^{-1}$, accompanied by the excitation of longitudinal electrostatic density waves. These waves arise from collective charge modulation driven by inter-stream interactions. As the free energy stored in the streaming beams is depleted, the wave amplitude decays, indicating the system's transition into the nonlinear regime. The continued decay of electrostatic energy destabilizes the BGK mode and further supports the interpretation of TSI as an inverse mechanism to Landau damping~\cite{Morgado_PRL_2017,Morse_PRL_1969}.
\begin{figure}[ht!]
    \centering
    \includegraphics[width=\columnwidth]{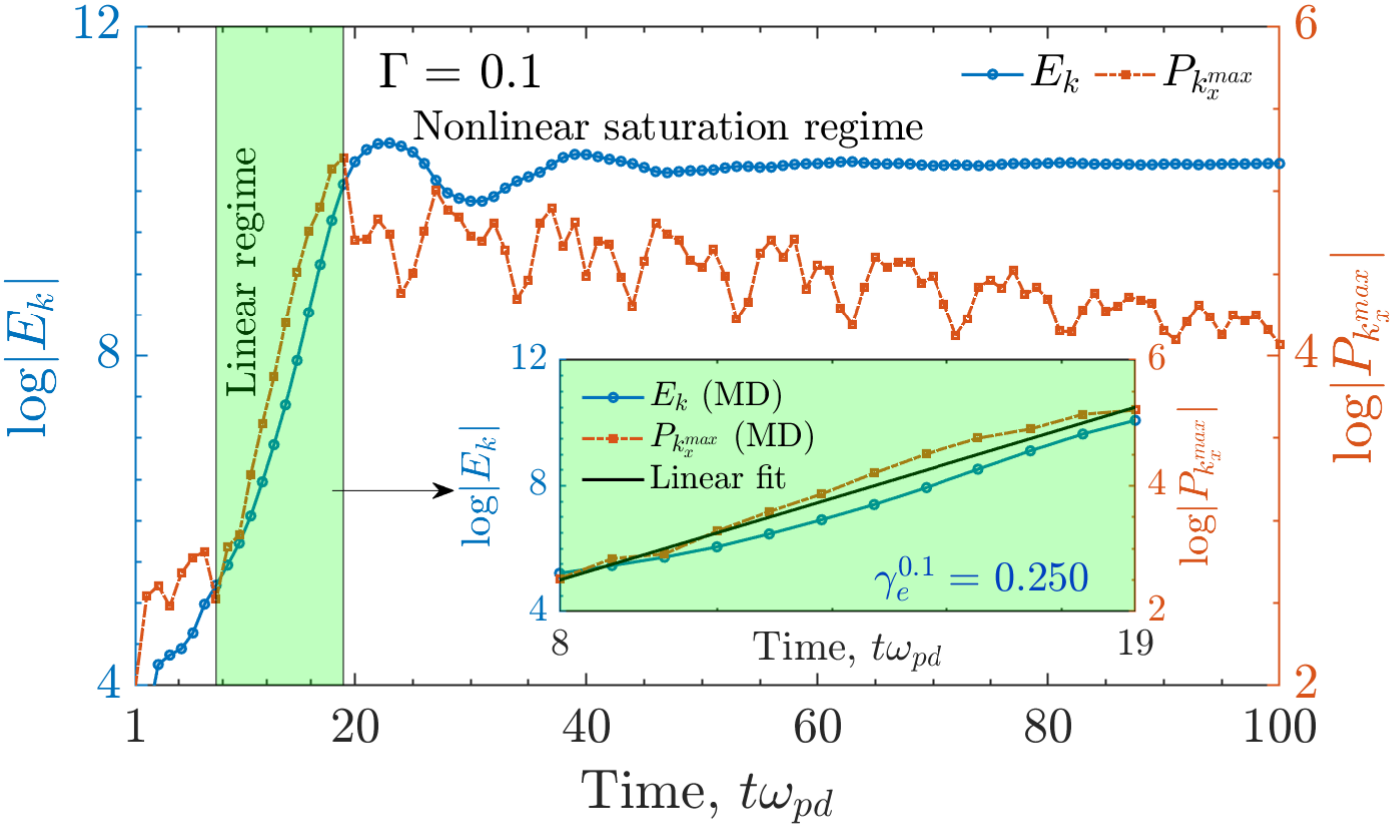}
    \caption{
    Evolution of the growth of TSI using perturbed kinetic energy $(|E_k|)$ (solid line) defined by Eq.~\eqref{Linear_KE_eqn}, and spectral energy $(|P_{k_x^{max}}|)$ (dash-dotted line) of the maximally growing mode. The inset shows TSI's linearly fitted growth rate in exponential regime $\gamma_e$ obtained from MD simulations.
    }
    \label{Figure_3}
\end{figure}
\begin{figure*}[ht!]
    \centering
    \includegraphics[width= \textwidth]{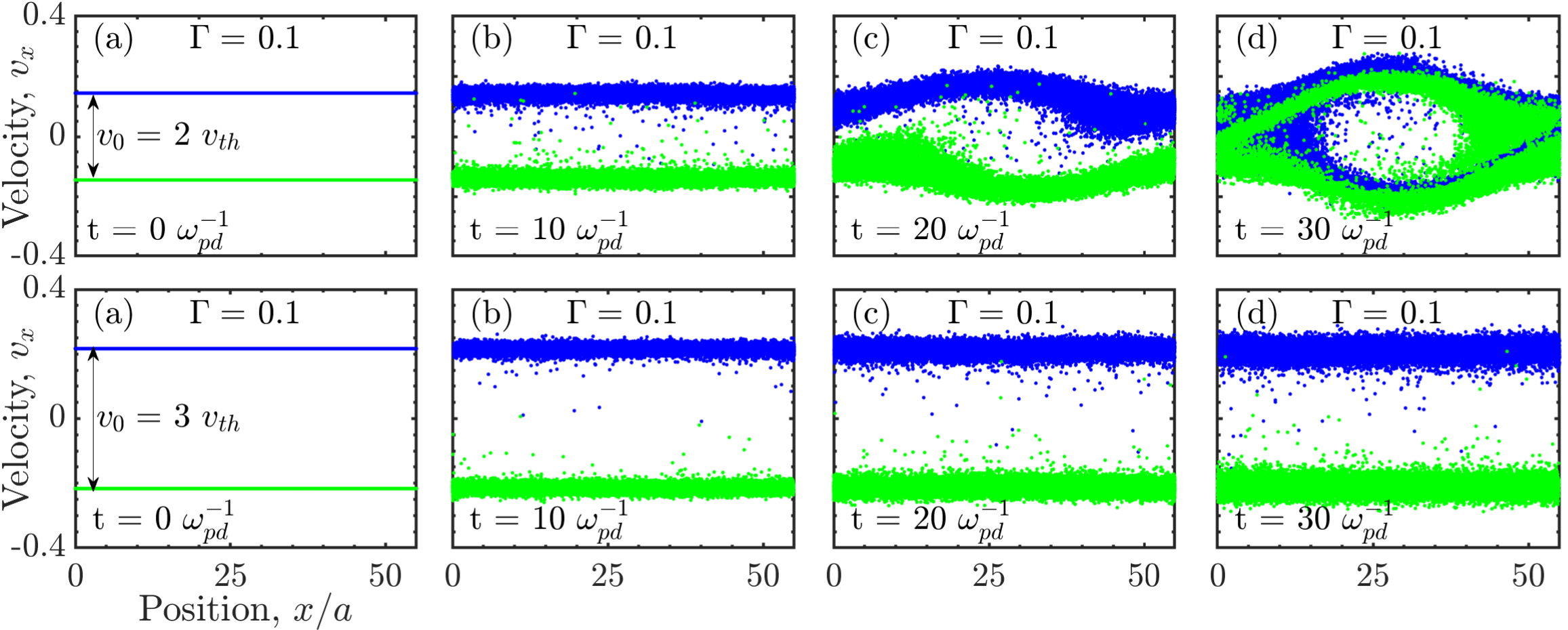}
    \caption{
    Effect of streaming velocity on the TSI for $\Gamma = 0.1$. 
    Top panel with $v_0 =2\ v_{th}$, and bottom panel with $v_0 = 3\ v_{th}$. The instability is more pronouncedly observed at $v_0 = 2\ v_{th}$, and completely vanishes for $v_0 = 3\ v_{th}$.
    }
    \label{Figure_4}
\end{figure*}
\subsection{Estimation of the linear growth rate}
\label{GR_estimation}
\paragraph*{}
The motive to determine the linear growth rate of an instability is vital to understand its early onset and timescale of instability, which offers a crucial benchmark for the validation of  theoretical predictions and simulations. To accurately quantify this growth, we employ two complementary approaches to measure the linear growth of TSI:
\begin{enumerate}
    \item 
    \textit{Perturbed kinetic energy method}: 
    We evaluate the perturbed kinetic energy of the system throughout its evolution by analyzing the MD simulation data. The perturbed kinetic energy is computed using the following expression:
\begin{equation}
    \label{Linear_KE_eqn}
    |E_k| = \sum_{i=1}^{N/2}  \left[ v_x^{i} \pm v_0 \right]^2 / \sum_{i=1}^{N/2} v_{0}^2.
\end{equation}
This method is widely employed to characterize the growth rate of both fluid and kinetic instabilities~\cite{Ashwin_PRL_2010, Pawandeep_POP_2021}. A similar root-mean-square velocity-based formulation has also been adopted in prior PIC simulations of the TSI~\cite{Ghorbanalilu_LPB_2014}.
    \item 
    \textit{Spectral energy method}:  
    In this approach, we extract the growth rate by tracking the evolution of spectral (wave) energy in the fastest growing mode. This is achieved through Fourier transformation of the spatial density histogram at successive time steps. The method has been previously validated in the analysis of both classical and quantum versions of the TSI~\cite{Williams_PRR_2019, Hu_PRE_2022}.
\end{enumerate}
\paragraph*{}
Figure~\ref{Figure_3} presents the linear growth of the instability as obtained from both methods. The solid line corresponds to the perturbed kinetic energy method described by Eq.~\eqref{Linear_KE_eqn}, while the dash-dotted line represents the spectral energy in the maximally growing mode. The inset shows the computed growth rate derived from the electrostatic wave spectrum. Notably, the growth rate obtained via the kinetic energy approach is approximately $1.40$ times higher than that derived from the spectral method. Despite this quantitative difference, both approaches exhibit consistent trends, affirming the reliability of the linear growth analysis during the early phase of the instability.
\paragraph*{}
The TSI is driven by the free energy associated with the excess kinetic energy of the counter-streaming beams. Its evolution can be broadly divided into two stages: the linear growth regime, where the available free energy is at its maximum, and the nonlinear saturation regime, where this energy is largely depleted. During the linear phase, the perturbed kinetic energy and the power associated with the maximally growing electrostatic mode exhibit similar exponential growths. In contrast, in the nonlinear saturation regime, their behaviors diverge: while the power in the dominant mode decays with irregular oscillations, the perturbed kinetic energy remains approximately constant. Initially, electrostatic waves grow by extracting free energy from the beams; however, once saturation is reached, the wave energy begins to decay steadily. The observed irregular oscillations in the wave spectrum are attributed to resonant energy exchange between particles and plasmons via wave–particle interactions. Similar oscillatory features have also been reported in quantum two-stream instabilities, where they arise from nonlinear wave–wave coupling mechanisms~\cite{Hu_PRE_2022}.
\begin{figure*}[ht!]
    \centering
    \includegraphics[width= \textwidth]{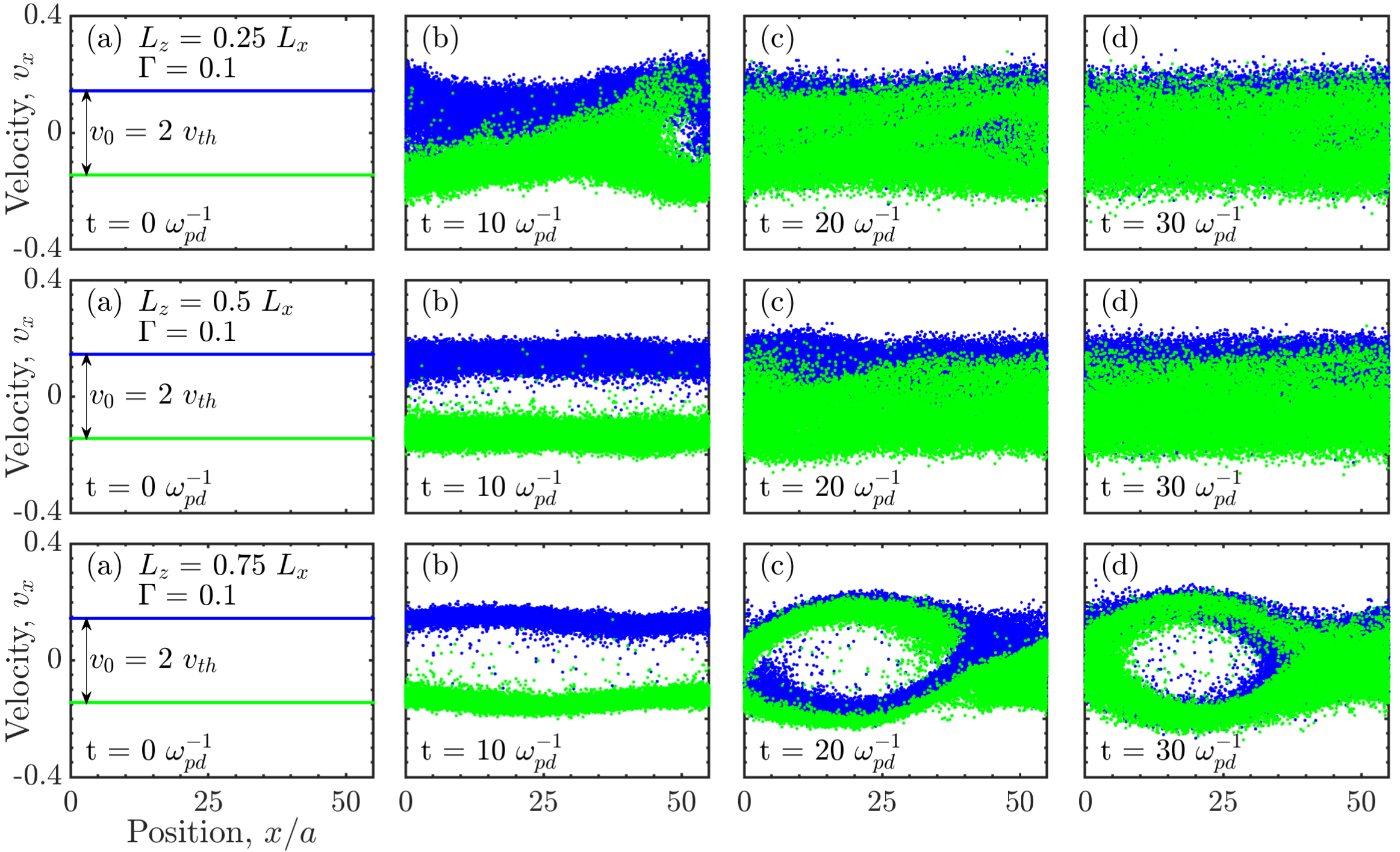}
    \caption{
    Effect of box size on the TSI for $\Gamma = 0.1$. Top panel with $L_z = 0.25 L_x$, middle panel with $L_z = 0.50 L_x$ and bottom panel with $L_z = 0.75 L_x$. However, in each panel, $L_x = L_y$. The increase in box dimension leads to a stabilized void formation in TSI. The bottom panel  of Fig.~\ref{Figure_6} shows the TSI for $\Gamma = 10$ when $L_x = L_y =L_z$.
    }
    \label{Figure_5}
\end{figure*}
\subsection{Effect of streaming velocity $v_0$ on TSI}
\label{TSI_MD_StreamEffect}
\paragraph*{}
\textcolor{black}{
The threshold value corresponds to the minimum relative streaming velocity required for the two-stream instability to overcome Landau damping and exhibit a positive growth rate. Physically, this condition arises when the free energy associated with the counter-streaming beams exceeds the damping due to resonant particles in the distribution function.
In our study, this threshold was empirically determined from the simulation results by identifying the lowest beam velocity at which a monotonic exponential growth in the electrostatic energy was observed. Below this velocity, perturbations are damped rather than amplified, indicating a stable regime. We also do not observe TSI for streaming velocity value higher than this threshold. }
The system supports TSI only when the velocity is close to the threshold value, i.e., $v_0 = 0.144~\mathrm{m/s}$. When the streaming velocity deviates significantly from this threshold in either direction, the electrostatic wave fails to grow. As a result, strong potential wells—required for particle trapping—do not form. Consequently, no BGK mode is triggered in the phase-space of the system.
\begin{figure*}[ht!]
    \centering
    \includegraphics[width= \textwidth]{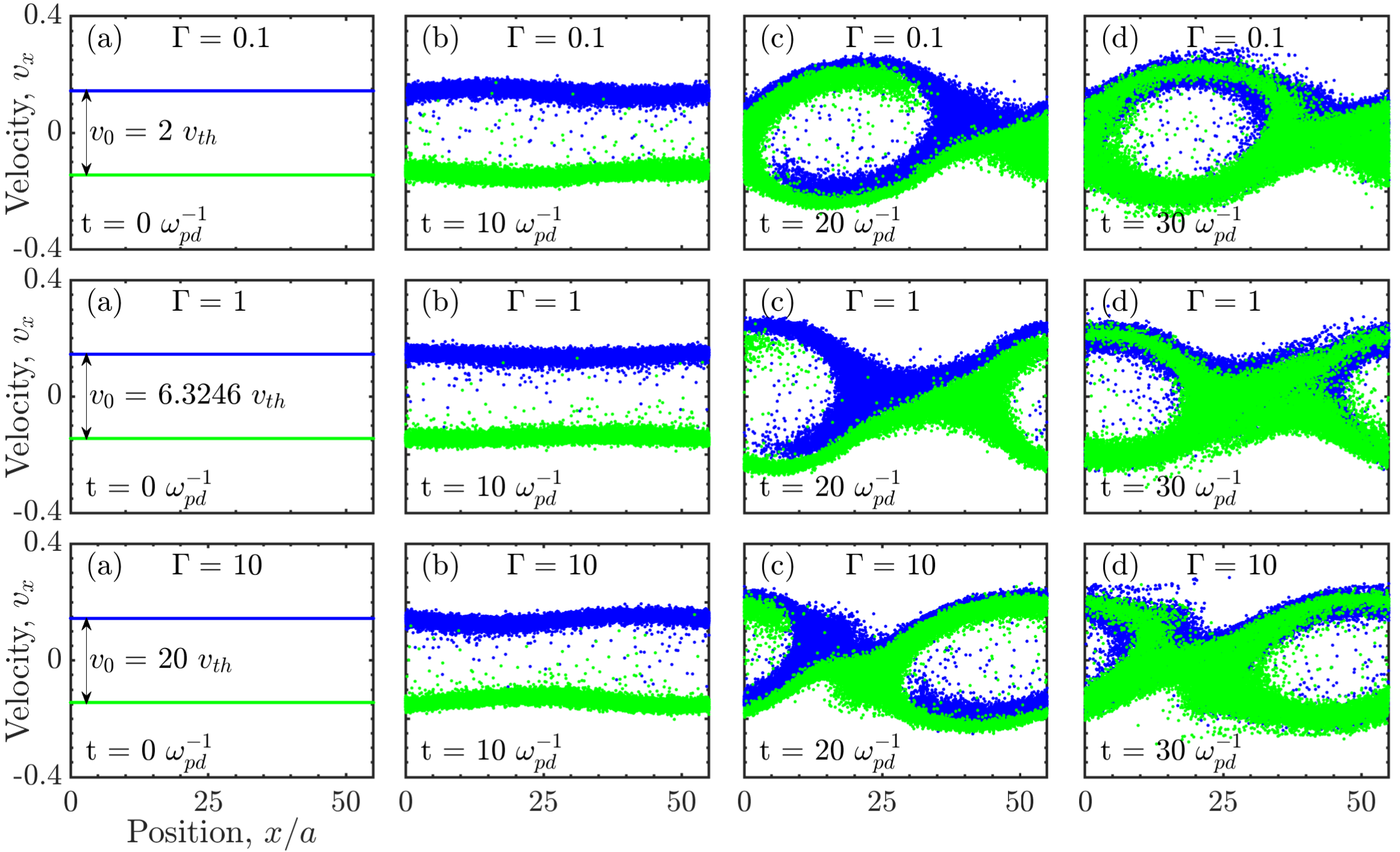}
    \caption{
    Phase-space evolution of the TSI at different coupling strengths with different streaming velocities in Coulomb OCP. The TSI is observed at $v_0 = 2\ v_{th}$, $6.3246\ v_{th}$, and $20\ v_{th}$ for $\Gamma = 0.1,\ 1,$ and $10$, respectively.
    }
    \label{Figure_6}
\end{figure*}
\paragraph*{}
Figure~\ref{Figure_4} illustrates the effect of streaming velocity on the TSI for a fixed coupling strength of $\Gamma = 0.1$. The instability fails to develop when the streaming velocity $v_0$ deviates significantly—either above or below—from the threshold value. The lower panels (a)–(d) in Fig.~\ref{Figure_4} clearly show the absence of BGK mode formation under such conditions. This indicates that the emergence of the BGK mode in TSI is strongly dependent on the streaming velocity of the particles.
\subsection{Effect of size of the simulation box on TSI}
\label{BoxSize_effect}
\paragraph*{}
An intriguing finding is that the geometry of the simulation domain strongly influences the formation of phase-space vortices. Specifically, the instability grows more rapidly when the box dimensions are unequal, while a distinct phase-space vortex still forms when the box is cubic with equal dimensions. This highlights how system dimensionality can impact the nonlinear evolution of the Two-Stream Instability.
\textcolor{black}{
We have done simulations to examine the effect of domain size aspect ratio on the instability development. We found that, for $L_z \geq L_x$, the instability grows at almost the same rate and with similar saturation behavior. It suggests that when a typical transverse threshold domain size $L_z = L_x$ is met, the instability is visible, and its dynamics generally follows trends similar to those of further increases in $L_z$.}
\paragraph*{}
Figure~\ref{Figure_5} shows the effect of box dimensions on TSI with strong coupling strength $\Gamma = 0.1$. The top panel shows the phase-space evolution of TSI with $L_z = 0.25L_x$. The system attains saturation very rapidly. The phase-space vortex is also not visible. The middle panel shows the phase-space evolution of TSI with $L_z = 0.50L_x$. The growth is a little slowed down compared to the earlier dimensions. The phase-space void is also starting to appear more clearly. The bottom panel shows the phase-space evolution of TSI with $L_z = 0.75L_x$. The vortex is visible, and the growth rate is also the slowest. The TSI in a system with equal box dimensions is shown in Fig.~\ref{Figure_1}.
\subsection{TSI at different strong coupling strengths}
\label{TSI_MD_StrongCoupling}
\paragraph*{}
\textcolor{black}{In the present study, the thermal velocity $v_{th}$ is determined from the coupling parameter $\Gamma$, which sets the temperature through Eq.~(2). For each chosen $\Gamma$ we varied the streaming velocity according to $v_0 = \alpha v_{th}$, where $\alpha$ is a dimensionless scaling factor known as streaming strength. Thus, the variation in $v_0$ across $\Gamma$ reflects changes in the ratio $v_0/v_{th}$, while the underlying coupling strength $\Gamma$ remains fixed. This approach ensures that the electrostatic and thermal energy balance inherent to a given $\Gamma$ is preserved, allowing us to isolate the effect of streaming strength on the growth of the instability.}
We successfully observe the TSI across all strong coupling regimes, provided the streaming velocity $v_0$ of the interpenetrating beams remains close to the threshold value, i.e., $v_0 = 0.144$ m/s. This confirms that the onset of TSI is robust within a well-defined velocity range. However, exploring the effect of strong coupling in our simulations is challenging because the imposed streaming velocity exceeds the thermal velocities corresponding to different coupling strengths, as summarized in Table~\ref{Table_2}. This results in a significant reduction of the effective coupling parameter $\Gamma$, effectively pushing the system into the weakly coupled regime. Consequently, while the TSI mechanism is clearly captured, the influence of strong coupling on its growth rate is masked by the dominance of kinetic effects introduced by the high streaming velocity.
\paragraph*{}
Figure~\ref{Figure_6} illustrates the development of the TSI across different strong coupling parameters. The top, middle, and bottom panels correspond to $\Gamma = 0.1$, $\Gamma = 1$, and $\Gamma = 10$, with respective streaming strengths $\alpha = 2$, $\alpha = 6.325$, and $\alpha = 20$. Notably, the instability exhibits the fastest growth for $\Gamma = 10$, suggesting a possible enhancement of the TSI in more strongly coupled regimes. However, it is important to recognize that at such high streaming strengths, the effective coupling parameter $\Gamma_{\text{eff}}$ may be significantly reduced due to heating effects arising from the initial kinetic energy imparted to the beams. This thermalization leads to a partial melting of the system, thereby diminishing the effective strong coupling.
\paragraph*{}
Moreover, as shown in Table~\ref{Table_2}, we observe a modest increase in the linear growth rate $\gamma_e$ with increasing $\Gamma$. While this trend appears consistent with an influence of strong coupling, we refrain from attributing the enhanced growth rate solely to $\Gamma$ due to the interplay of additional kinetic effects inherent to the simulation dynamics. These findings highlight the complexity of isolating pure strong coupling effects in the presence of high streaming velocities and open avenues for further investigations under more controlled conditions.
\begin{table}[!ht]
\caption{Growthrate $\gamma_e$ of TSI as a function of streaming strength $\alpha$ for different coupling strength $\Gamma$.}
{\renewcommand{\arraystretch}{2}
\begin{tabular}{|l|l|l|l|l|l|}
\hline  
\textbf{Coulomb coupling,   $\Gamma$}  & 0.01   &  0.1  & 1      & 10     & 100   \\ \hline 
\textbf{Streaming strength, $\alpha$}  & 0.6325 & 2     & 6.325  & 20     & 63.25 \\ \hline 
\textbf{Growthrate, $\gamma_e$}        & 0.25   & 0.29  & 0.29   & 0.31   & 0.31  \\\hline 
\end{tabular}
}
\label{Table_2}
\end{table}
\subsection{Spectral analysis}
\label{TSI_MD_Spectral}
\begin{figure}[ht!]
    \centering
  \includegraphics[width=\columnwidth]{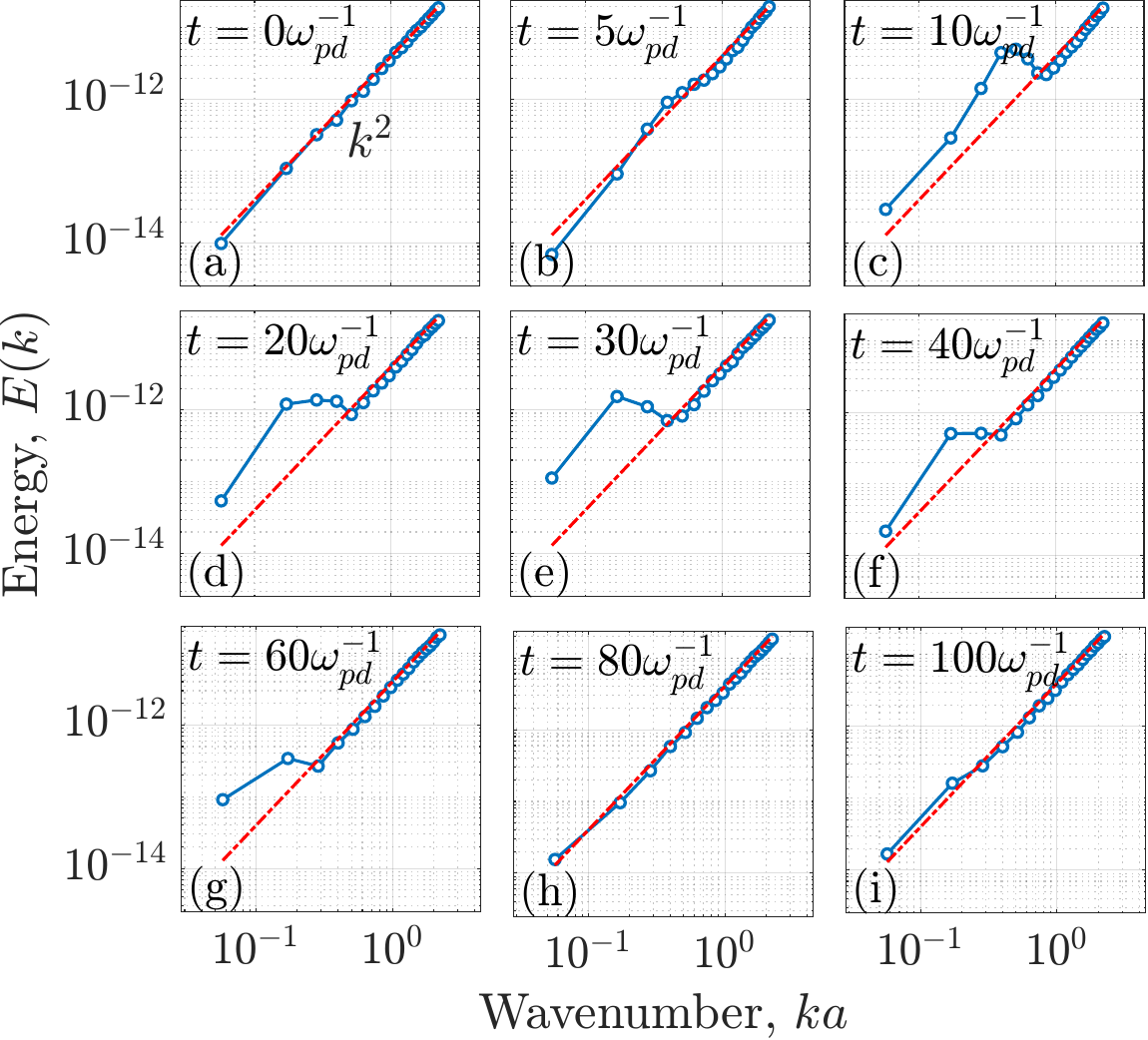}
    \caption{
    Time evolution of the 3D power spectrum $E(k)$ for the TSI. Initially (a–b), the spectrum shows thermalized $k^2$ scale (thermal equilibrium of the system). As time progresses (c–f), the BGK mode emerges, the power spectrum shows the hump at that particular scale. At later times (g–i), the modes again dump energy to the particles which results in thermalization.}
    \label{Figure_7}
\end{figure}

\paragraph*{}
The TSI is a kinetic instability characterized by the transfer of energy from particle streaming motions to collective wave modes, which subsequently grow exponentially. In the present model, the electrostatic wave dynamics evolve on dust timescales. Initially, the wave energy is broadly distributed across particle-level fluctuations. As the instability develops, energy is progressively funneled into intermediate and large-scale modes, extracted from the kinetic energy of the particles. To visualize this modal energy transfer, the temporal evolution of the energy spectrum $E(k)$ is shown in Fig.~\ref{Figure_7}, capturing the redistribution of energy in wave-number space throughout the instability's growth and saturation phases.
\paragraph*{}
The energy spectrum is computed from the coarse-grained velocity  field $\mathbf{u}(\mathbf{x,y,z}, t)$. A three-dimensional Fourier transform is applied to obtain the velocity field  $\mathbf{u}(\mathbf{k}, t)$, in wave-number space, from which the modal energy is calculated as $E(\mathbf{k}, t) = |\mathbf{u}(\mathbf{k}, t)|^2 / 2$. To obtain the one-dimensional shell-averaged energy spectrum, $E(k, t)$, we integrate over all modes within spherical shells of unit width in $k$-space, defined by $k - 1 \leq |\mathbf{k}'| \leq k$. The resulting spectrum is given by:
\begin{equation}
    E(k,t) = \sum_{k - 1 \leq |\mathbf{k}'| \leq k} E(\mathbf{k}', t).
\end{equation}
\paragraph*{}
Initially, the energy spectrum exhibits a characteristic $k^2$ scaling across the Fourier modes, indicative of thermalization~\cite{Rauoof_PLA_2024}. This behavior persists despite the injection of energy through two counter-streaming beams. Since the particles comprising each stream are randomly distributed throughout the simulation domain, the injected energy is rapidly redistributed among all fluctuation modes. As a result, the initial energy input does not perturb the thermal equilibrium scaling, and the $k^2$ spectral signature remains preserved during the early stages.
\paragraph*{}
However, as the BGK mode begins to develop, a distinct hump emerges in the power spectrum at a particular wavenumber, as seen in Fig.~\ref{Figure_7}(c). At later times—Fig.~\ref{Figure_7}(d) to (f)—the spectral energy shifts toward a lower wavenumber, corresponding to a spatial mode of twice the original scale. This shift is a signature of hole merger, a well-established nonlinear phenomenon in TSI. However, this merger process is not clearly discernible in the phase-space trajectories shown in Fig.~\ref{Figure_1}. Around $t \approx 60~\omega_{pd}^{-1}$, the dominant mode saturates, and the energy spectrum gradually returns to a thermalized $k^2$ scaling, indicating that the BGK mode contributes significantly to the system’s progression towards equilibrium. Additionally, the spectral hump exhibits persistent oscillations over time, with a characteristic period of $2$–$3~\omega_{pd}^{-1}$, which corresponds to the rotation frequency of the phase-space hole, thereby highlighting the connection between wave dynamics and phase-space structures.
\begin{figure}[ht!]
    \centering
  \includegraphics[width=\columnwidth]{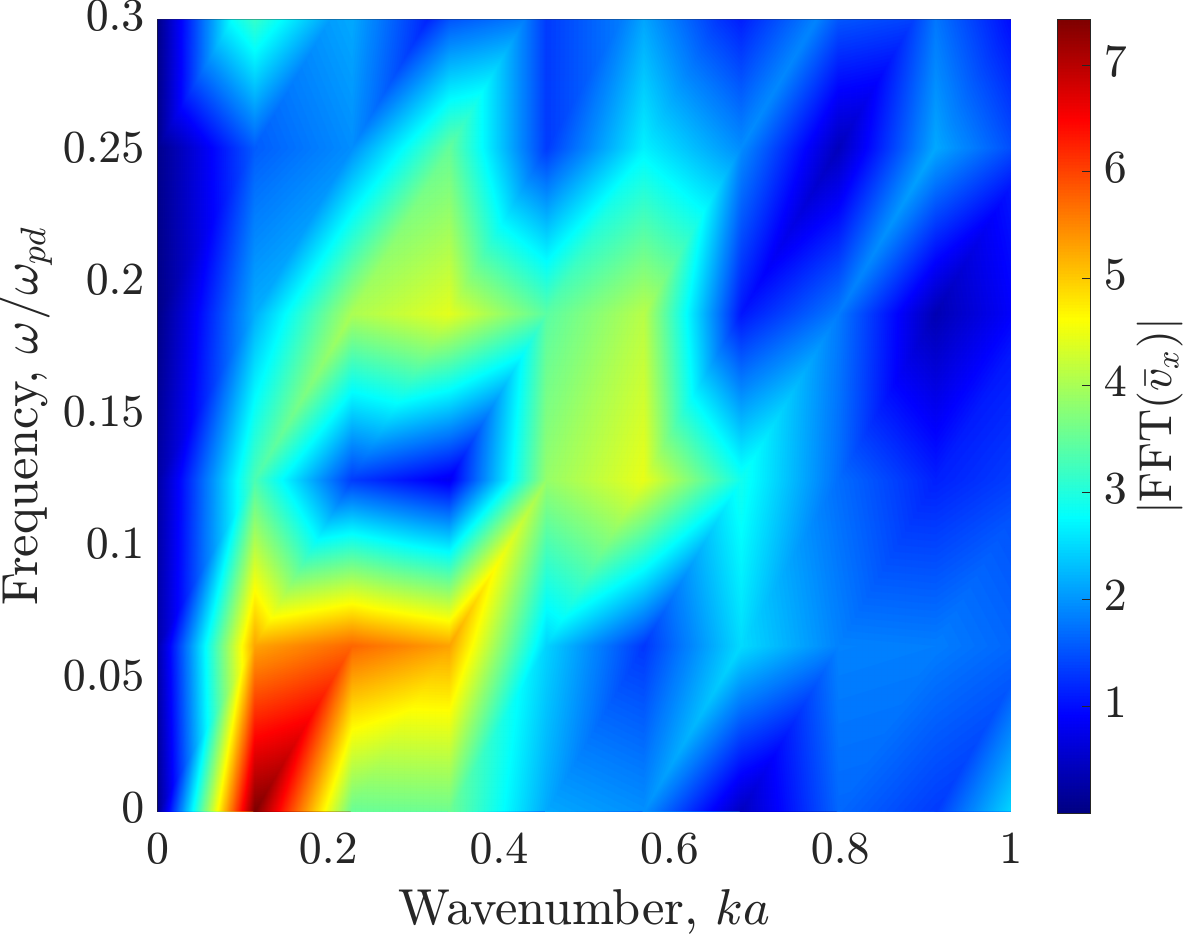}
    \caption{Dispersion relation for TSI obtained from fluidized data based on the MD simulations. The wavenumber $k$ is normalized by the average interparticle separation $a$ and the frequency $\omega$ is normalized by dust plasma frequency $\omega_{pd}$. }
    \label{Figure_8}
\end{figure}
\paragraph*{}
To further elucidate the modal structure and identify the dominant wave modes generated during the nonlinear evolution of the TSI, we compute the dispersion relation—i.e., the relationship between wave frequency $\omega$ and wavevector $k$—by performing a Fourier transform in both space and time on the mean velocity field $v_x$. To isolate the relevant longitudinal modes, we first obtain the coarse-grained $v_x$ and integrate it along the transverse directions ($y$ and $z$). The resulting $\omega$–$k$ plot, shown in Fig.~\ref{Figure_8}, clearly exhibits a bright peak indicating the presence of a coherent wave mode. This dominant mode approximately corresponds to the spectral hump seen at later times in Fig.~\ref{Figure_7}, confirming the emergence of a BGK-like structure. Notably, the observed dispersion is Doppler-shifted due to the relative motion between the particles and the wave frame. In the wave frame, the BGK structure appears stationary, but in the laboratory frame, the finite drift velocity of the streams causes a shift in frequency proportional to $k v_0$, where $v_0$ is the drift speed of the background flow. This Doppler shift is a crucial diagnostic that helps distinguish between true wave growth and a mere advection of fluctuations, and its presence here supports the interpretation that a nonlinear BGK mode is responsible for the observed spectral features.
\section{Summary and Conclusions}
\label{Sum_Con}
\paragraph*{}
In this study, we have explored the fascinating dynamics of the Two-Stream Instability in a one-component plasma, where the pair-wise particle interactions are governed by a long-range Coulomb potential, using fully three-dimensional classical molecular dynamics simulations. By simulating counter-streaming charged particle beams interacting via long-range Coulomb forces, we have successfully observed the hallmark features of TSI, namely,  phase-space vortex formation, charge bunching, and electrostatic density perturbations. These perturbations emerge as the system taps into the free energy available in the streaming particles, and eventually damp out as the system evolves toward a new equilibrium. A key finding is that TSI manifests robustly across all coupling strengths, provided the streaming velocity lies close to a critical threshold, specifically at $v_0 = 0.144$ m/s. This highlights the universality of TSI onset in such systems when proper conditions are met. Interestingly, in fully 3D geometries with equal dimensions, the instability exhibits a slower and smoother growth—suggesting subtle geometric effects that moderate the nonlinear evolution of the instability.
\paragraph*{}
The emergence of the Bernstein–Greene–Kruskal mode further confirms the onset of TSI. These nonlinear structures capture the interaction between the two beams in phase space and provide deep insights into the kinetic processes underlying the instability. While single BGK vortices were found to be energetically unstable and gradually dissipated, multiple BGK modes showed a tendency to merge into a dominant mode—a dynamic that remains intriguing and rich for future kinetic investigations. Importantly, our results suggest that the velocity separation between the beams strongly influences the nature and dynamics of vortex formation.
\paragraph*{}
\textcolor{black}{
Our work opens several exciting directions for future exploration. Most notably, we observed that introducing a shielded Coulomb interaction—even with minimal screening ($\kappa = 0.01$)—can suppress the TSI entirely, calling for a deeper theoretical understanding of the conditions necessary for instability in strongly coupled plasmas. Moreover, the role of initial velocity perturbations and their mode structures in triggering TSI remains an open challenge, one that demands detailed kinetic modeling of the dispersion relations involved.
Finally, the influence of strong coupling remains an area ripe for discovery. While our simulations reveal that higher $\Gamma$ values correlate with increased instability growth, the high streaming velocities employed likely modify the effective coupling strength. The impact of strong coupling $\Gamma$ necessitates a novel theoretical framework based on the substantial disparities between the streaming and thermal velocities observed in current studies. A more refined theoretical framework is needed to fully disentangle the effects of strong coupling on TSI growth, vortex formation, and coalescence of eddies. These questions not only enhance our understanding of kinetic instabilities in strongly coupled plasmas but also pave the way for novel approaches in plasma control and diagnostics in both laboratory and astrophysical settings.}
\begin{acknowledgements}
Work done by AM and ST was supported by the Indian Institute of Technology Jammu Seed Grant No. SG0012. 
AM and ST acknowledge the use of AGASTYA High-Performance Computing for present studies. 
ST also acknowledges the Science and Engineering Research Board (SERB) Grant No. CRG/2020/003653 for partial support for the work. 
AS is grateful to the Indian National Science Academy (INSA)
for the INSA Honorary Scientist position.
\end{acknowledgements}
\bibliography{TSI_Plasma}
\end{document}